\documentclass[iop,apj,twocolappendix]{emulateapj}
\usepackage[colorlinks, linkcolor=red, citecolor=blue, linktocpage=true]{hyperref}

\begin{document}

\title{Detection of three Gamma-Ray Burst host galaxies at $z\sim6$}

\author{J. T. W. M\textsuperscript{c}Guire\altaffilmark{1}, N. R. Tanvir\altaffilmark{1}, 
	A. J. Levan\altaffilmark{2}, M. Trenti\altaffilmark{3}, E. R. Stanway\altaffilmark{2},
	J. M. Shull\altaffilmark{4}, K. Wiersema\altaffilmark{1}, D. A. Perley\altaffilmark{5},
	R. L. C. Starling\altaffilmark{1}, M. Bremer\altaffilmark{6}, J. T. Stocke\altaffilmark{4},
	J. Hjorth\altaffilmark{5}, J. E. Rhoads\altaffilmark{7}, E. Curtis-Lake\altaffilmark{8}, 
	S. Schulze\altaffilmark{9,10}, E. M. Levesque\altaffilmark{11}, B. Robertson\altaffilmark{12}, 
	J. P. U. Fynbo\altaffilmark{5}, R. S. Ellis\altaffilmark{13}, 
	A. S. Fruchter\altaffilmark{14}}

\altaffiltext{1}{Department of Physics \& Astronomy, University of Leicester, University Road, Leicester LE1 7RH, UK; jtwm1@le.ac.uk}
\altaffiltext{2}{Department of Physics, University of Warwick, Gibbet Hill Road, Coventry CV4 7AL, UK}
\altaffiltext{3}{School of Physics, The University of Melbourne, VIC 3010, Australia}
\altaffiltext{4}{CASA, Department of Astrophysical \& Planetary Sciences, University of Colorado, Boulder, CO 80309, USA}
\altaffiltext{5}{Dark Cosmology Centre, Niels Bohr Institute, Copenhagen University, Juliane Maries Vej 30, 2100 K\o benhavn \O, Denmark}
\altaffiltext{6}{H H Wills Physics Laboratory, University of Bristol, Tyndall Avenue, Bristol BS8 1TL, UK}
\altaffiltext{7}{School of Earth and Space Exploration, Arizona State University, Tempe, AZ 85287, USA}
\altaffiltext{8}{Scottish Universities Physics Alliance (SUPA), Institute for Astronomy, University of Edinburgh, Royal Observatory, Edinburgh EH9 3HJ, UK}
\altaffiltext{9}{Instituto de Astrof\'isica, Facultad de F\'isica, Pontificia Universidad Cat\'olica de Chile, Vicu\~{n}a Mackenna 4860, 7820436 Macul, Santiago, Chile}
\altaffiltext{10}{Millennium Institute of Astrophysics, Vicu\~{n}a Mackenna 4860, 7820436 Macul, Santiago, Chile}
\altaffiltext{11}{Department of Astronomy, Box 351580, University of Washington, Seattle, WA 98195, USA}
\altaffiltext{12}{Department of Astronomy and Astrophysics, University of California, Santa Cruz, CA 95064, USA}
\altaffiltext{13}{European Southern Observatory, Karl Schwarzschild Strasse 2, D-85386 Garching, Germany}
\altaffiltext{14}{Space Telescope Science Institute, Baltimore, MD 21218, USA}

\begin{abstract}
Long-duration Gamma-Ray Bursts (GRBs) allow us to pinpoint and study star-forming galaxies in the early universe, thanks to their orders of magnitude brighter peak luminosities compared to other astrophysical sources, and their association with deaths of massive stars. We present {\em Hubble Space Telescope} Wide Field Camera 3 detections of three {\em Swift} GRB host galaxies lying at redshifts $z = 5.913$ (GRB 130606A), $z = 6.295$ (GRB 050904), and $z = 6.327$ (GRB 140515A) in the F140W (wide-$JH$ band, $\lambda_{\rm{obs}}\sim1.4\,\micron$) filter. The hosts have magnitudes (corrected for Galactic extinction) of $m_{\rm{\lambda_{obs},AB}}= 26.34^{+0.14}_{-0.16}, 27.56^{+0.18}_{-0.22},$ and $28.30^{+0.25}_{-0.33}$ respectively. In all three cases the probability of chance coincidence of lower redshift galaxies is $\la2\,\%$, indicating that the detected galaxies are most likely the GRB hosts. These are the first detections of high redshift ($z > 5$) GRB host galaxies in emission. The galaxies have luminosities in the range $0.1-0.6\,L^{*}_{z=6}$ (with $M_{1600}^{*}=-20.95\pm0.12$), and half-light radii in the range $0.6-0.9\,\rm{kpc}$. Both their half-light radii and luminosities are consistent with existing samples of Lyman-break galaxies at $z\sim6$. Spectroscopic analysis of the GRB afterglows indicate low metallicities ($[\rm{M/H}]\la-1$) and low dust extinction ($A_{\rm{V}}\la0.1$) along the line of sight. Using stellar population synthesis models, we explore the implications of each galaxy's luminosity for its possible star formation history, and consider the potential for emission-line metallicity determination with the upcoming {\em James Webb Space Telescope}.
\end{abstract}


\section{Introduction}
Long-duration Gamma-Ray Bursts (GRBs) are associated with the core collapse of very massive stars. At peak, their optical afterglows can be orders of magnitude brighter than the next most luminous astrophysical sources \citep[e.g.,][]{rac08,blo09}. Spanning the majority of cosmological time, GRBs found by the {\em Swift} satellite have been detected from $z\sim0.03$ \citep{pia06} to $z=8-9$ \citep{tan09,sal09,cuc11} with a median redshift of $z\sim2$ \citep{jak06,fyn09,hjo12}. GRBs, then, are cosmological probes, sampling sightlines through individual galaxies \citep[e.g.,][]{jak04,dup12} and giving us insight, via afterglow spectroscopy, into a large array of local (galactic) and intergalactic properties, such as metal abundance \citep[e.g.,][]{cuc15}, temperature and gas densities \citep[e.g.,][]{sta13,cam15}, dust content \citep[e.g.,][]{zaf11,scha12}, and also the neutral fractions of the intergalactic medium (IGM) \citep[e.g.,][]{tot06,har15}.

GRBs are also important because they select star-forming hosts independently of the luminosity of the galaxies themselves. In some cases the hosts are bright enough for further follow-up, allowing comparison of their properties in emission with those in absorption \citep[e.g.,][]{kru15}. However, at higher redshifts it becomes increasingly challenging to detect the hosts directly. In fact, this trend can be turned to an advantage since it means that the ratio of undetected to detected hosts in deep imaging provides a measure of the proportion of star formation occurring in very faint galaxies beyond the depths of conventional flux-limited galaxy surveys. Until now, no hosts have been detected in emission beyond $z\sim5$ \citep{cha07}, which is consistent with a steep faint-end slope of the galaxy ultraviolet (UV) luminosity function at high redshifts \citep{tan12,tre12,bas12}.

The majority of galaxies known from early cosmic times have been identified via the Lyman-break technique \citep{ste96}, in which high redshift candidates are selected by their presence in images with red (typically near infrared) passbands, along with their absence in images with bluer passbands, taken in (typically optical) ``veto" filters, presumed to be shortward of the Ly$\alpha$ break in the rest frame. This has resulted in samples of up to $\sim1000$ Lyman-break galaxies (LBGs) with photometric redshifts around $z\sim6$, though there has been recent work in pushing the envelope to $z\sim9-10$ \citep[e.g.,][]{oes14}. These are primarily found through various {\em Hubble Space Telescope} ({\em HST}) deep imaging campaigns \citep[e.g.,][]{bou15,dun15,oes15a}. Our understanding of the physical properties of these galaxies, however, has largely been limited to what can be learnt from their broad-band colours \citep[e.g.,][]{wil13}, and spectroscopy of a handful of intrinsically very luminous or highly lensed examples \citep[e.g.,][]{oes15b,stark15}. The very blue UV continua measured for some faint $z\sim7$ LBG samples led to suggestions that these galaxies must be both dust free and have very low metallicity \citep[$\la1\%\,Z_{\odot}$;][]{bou10}, although subsequent analyses have found rather less extreme colours \citep{dun12,fin12}. Early Atacama Large Millimetre Array (ALMA) studies of the continuum and [\ion{C}{2}] ($158\,\micron$) line emission properties of small samples of $z\sim6$ LBGs are also consistent with their being low-dust and moderately low-metallicity systems \citep{cap15}.

There is now strong evidence that GRBs preferentially occur in low-metallicity, star-forming galaxies \citep[e.g.,][]{lev10,gra13,cuc15,gra15a}, although a small fraction are found in high-metallicity environments \citep[e.g.,][]{gra15b}. From this, we would expect that the brightest high-$z$ GRB hosts would satisfy the selection criteria of current LBG surveys \citep[e.g.,][]{fyn08}, and hence, through the afterglow spectroscopy, potentially provide evidence of the internal conditions in LBGs, which are otherwise poorly constrained.

In this paper we report on the detection and properties of the host galaxies of three GRBs, 130606A, 050904, and 140515A, at spectroscopic afterglow redshifts of $z=5.913, z=6.295$, and $z=6.327$ respectively. These are the three most distant GRB hosts directly detected to date. We first discuss the GRB sample, and their host properties inferred from their afterglows in Section~\ref{sec:2}. The {\em HST} observations, and data analysis methods are discussed in Section~\ref{sec:3}. In Section~\ref{sec:4} we demonstrate that these galaxies are unlikely to be low-$z$ interlopers. Assuming they are the hosts, we compare them to LBGs and the lower redshift population of GRB hosts. Finally in Section~\ref{sec:5}, we briefly discuss the implications of our results, along with the potential of {\em James Webb Space Telescope} observations of these galaxies. Details of the relative astrometry procedure are given in Appendix~\ref{apsec:1}, which determines the precision with which we can locate the afterglow positions on our {\em HST} images. A statistical study of the high redshift GRB host sample, including an updated analysis of the non-detections, will be presented in a future paper. That work will address the implications of the host luminosities for the faint end of the galaxy luminosity function (LF) at $z>6$, which is of key importance for our understanding of reionization. 

We assume a $\Lambda$-CDM cosmology using the new 2015 {\em Planck} results \citep{pck15}, with $\Omega_{\rm{M}}=0.308$, $\Omega_{\Lambda}=0.692$, and $H_{0}=67.8\,\rm{km}\,\rm{s}^{-1}\,\rm{Mpc}^{-1}$. AB magnitudes \citep{oke83} and uncertainties at the $1\sigma$ confidence level are presented throughout unless otherwise stated. 

\section{GRB Sample}
\label{sec:2}
\subsection{GRB 130606A}
\label{sub:2.1}
GRB 130606A was confirmed as a high-redshift source from optical spectroscopic observations with the Gran Telescopio Canarias $10.4\,\rm{m}$ telescope, giving $z=5.913$ \citep{cas13}. X-shooter observations by \citet{har15}, using the European Southern Observatory (ESO) Very Large Telescope (VLT), found a neutral hydrogen column density of $\log[N_{\rm{HI}}\,(\rm{cm}^{-2})]=19.91\pm0.02$ based on the red damping wing of the Ly$\alpha$ absorption line, consistent with results from spectroscopy of the event by other groups \citep{cho13,tot14}. \citet{har15} estimated the average metallicity of the host galaxy to be $-1.7<[\rm{M/H}]<-0.9$, while multiple intervening absorption lines were detected at redshifts across the range $z=2.52$ to $z=4.65$, with a possible intervening system at $z=5.806$. They also found host dust extinction along the line of sight, measured using the X-ray and optical spectral energy distribution (SED), to be consistent with zero and with a $3\sigma$ upper limit of $A_{\rm{V}}<0.2\,\rm{mag}$.

\subsection{GRB 050904}
\label{sub:2.2}
GRB 050904 was confirmed as a high-redshift source from optical spectroscopic observations with the Subaru $8.2\,\rm{m}$ Telescope, giving $z=6.295$ \citep{kaw06}. A detailed analysis was conducted by \citet{tot06}, who measured a neutral hydrogen column density of $\log[N_{\rm{HI}}\,(\rm{cm}^{-2})]\approx21.6$ from the damped Ly$\alpha$ system associated with the host galaxy, and detected an intervening absorber at $z=4.840$. \citet{tho13} re-analyzed the Subaru spectrum and estimated a metallicity of $[\rm{M/H}]\sim-1.6\pm0.3$, based solely on the \ion{S}{2} ($\lambda1253$) line (the other metal lines being saturated and/or blended). Considering the low resolution of the spectrum, this estimate should probably be regarded as a lower limit. Host extinction was determined to be $A_{\rm{V}}=0.01\pm0.02\,\rm{mag}$ \citep{zaf10}.

A previous search for the host galaxy was conducted by \citet{ber07}, with {\em HST} and {\em Spitzer} observations. They observed with both {\em HST's} Advanced Camera for Surveys (ACS), and Near Infrared Camera and Multi-Object Spectrometer (NICMOS), and their F850LP and F160W filters respectively. {\em Spitzer} observations were carried out with the Infrared Array Camera (IRAC) in all four channels ($3.6, 4.5, 5.8, \rm{and}\,8.0\,\micron$). The host galaxy was undetected down to $3\sigma$ upper limits of $m_{\rm{F160W}}>27.2\,\rm{mag}$ and $m_{3.6\micron}>25.35\,\rm{mag}$ \citep[see also][]{per15}. The ACS data were re-reduced \citep{tan12}, and a $3\sigma$ upper limit of $m_{\rm{F850LP}}>26.36\,\rm{mag}$ was estimated, accounting for flux loss from IGM absorption. 

\subsection{GRB 140515A}
\label{sub:2.3}
GRB 140515A was confirmed as a high-redshift source from optical spectroscopic observations with the Gemini-North $8.1\,\rm{m}$ Telescope, giving $z=6.327$ \citep{cho14}. \citet{mel15} presented ESO/VLT X-shooter observations of the afterglow, estimating a neutral hydrogen column density of $\log[N_{\rm{HI}}\,(\rm{cm}^{-2})]\la18.5$ from the red damping wing of the Ly$\alpha$ absorption line. They determined $3\sigma$ upper limits for host galaxy metallicity, measuring $[\rm{Si/H}]<-1.4$, $[\rm{O/H}]<-1.1$ and $[\rm{C/H}]<-1.0$. An intervening absorber was detected at $z=4.804$, and they also determined host extinction to be $A_{\rm{V}}\approx0.1\,\rm{mag}$. In addition, \citet{mel15} point out that the low hydrogen column would be consistent with GRB 140515A having exploded in a relatively low-density galactic environment.

\section{Observations and Results}
\label{sec:3}
\begin{figure*}
	\epsscale{1.0}
	\plotone{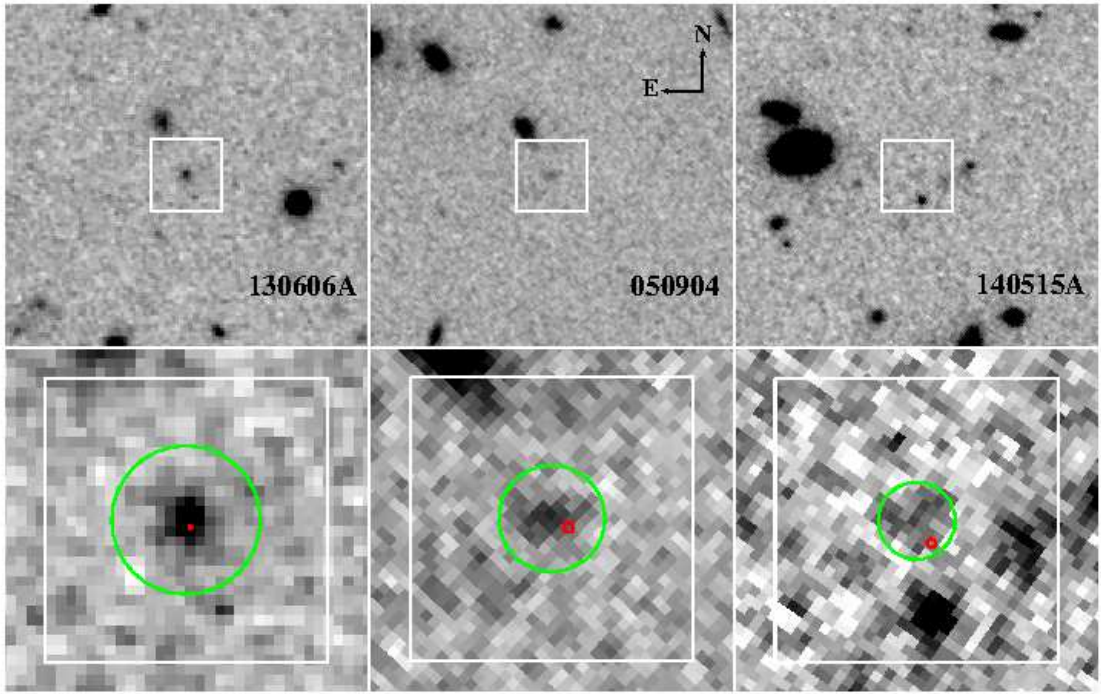}
	\caption{Observation field cut-outs for each GRB (upper panels) with a zoom-in at higher contrast for each detected galaxy (lower panels). GRB afterglow positional uncertainty at $1\sigma$ is shown as red circles, while the detected galaxies are encircled with their customized aperture in green (see Section~\ref{sub:3.1}). The white box surrounding each galaxy has sides of length $2\arcsec$ and is for scaling purposes.\label{fig:1}}
\end{figure*} 

We observed the GRB positions with {\em HST} as part of a program to study the host galaxies of $z\ga6$ GRBs (GO-13831; PI: Tanvir). Observations were performed with the Wide Field Camera 3 Infrared Channel (WFC3-IR) using the F140W filter. Spanning the conventional $J$- and $H$-bands, this filter has a pivot wavelength of $\lambda_{\rm{obs}}=13920\,\rm{\AA}$, and bandpass of $\Delta\lambda=3840\,\rm{\AA}$. The F140W filter was chosen due to its wide bandpass so as to maximise sensitivity in the rest-frame UV, while also avoiding bright nebular emission lines which can otherwise add significant uncertainties to SED modelling \citep[e.g.,][]{scha09,gon12,gon14,oes15b}.

A three-point dither was adopted within each orbit with further shifts between orbits to provide an optimal six-point dither pattern, while simultaneously stepping over WFC3-IR detector ``blobs". A log of the observations is summarized in Table~\ref{tb:1}. We processed the data using {\tt AstroDrizzle}\footnote{\url{http://drizzlepac.stsci.edu/}} resulting in a pixel scale of $0\farcs07\,\rm{pixel}^{-1}$, which is half the native detector pixel size, for all GRB fields.

Relative astrometry, locating the GRBs precisely on the {\em HST}/WFC3 images, was achieved by directly comparing the positions of objects (mostly stars) on the {\em HST} images with their positions on images showing the afterglows (see Appendix~\ref{apsec:1} for greater detail). The precision with which this can be done depends on the number and brightness of the comparison sources and also the $S/N$ of the afterglow detections. In all three cases, the $1\sigma$ positional uncertainty is $<0\farcs05$.

For GRBs 130606A and 140515A, we fixed the absolute astrometry using the reported radio positions for the afterglows \citep{las13,las14}. For GRB 050904 the absolute astrometry was tied to SDSS stars visible in our ground-based comparison images \citep{pie03}.

In all three cases, the observations reveal galaxies coincident within the inferred GRB afterglow locations. As discussed in Section~\ref{sub:4.1}, we find the probability that these galaxies are chance alignments is small and believe they are likely to be the GRB hosts. Indeed, we note that typically GRBs lie on the UV-bright regions of their hosts \citep{fru06,sve10}, consistent with what is seen here. Cutouts of the fields are shown in Figure~\ref{fig:1}, with the GRB positions shown as red circles, and the galaxies within green circles. The galaxy centroid positions (accurate to $\la0\farcs1$) and offsets from GRB afterglow locations are given in Table~\ref{tb:2}.

\begin{deluxetable*}{lccc}
	\tablecolumns{4}
	\tablewidth{0pt}
	\tablecaption{Log of HST observations.\label{tb:1}}
	\tablehead{
	\colhead{Identifier} & \colhead{130606A} & \colhead{050904} & \colhead{140515A}
	}
	\startdata
	Date                              & 2015 Aug 13 & 2014 Oct 31 & 2015 Feb 01 \\
	UT Time                           & 02:29:36    & 11:15:51    & 15:06:38    \\
	$\lambda_{\rm{rest}}\,(\rm{\AA})$ & 2014        & 1908        & 1900        \\
	Exposure (s)                      & 10791       & 13488       & 10791       \\
	Redshift\tablenotemark{a}         & 5.913       & 6.295       & 6.327       \\
	$A_{\rm{F140W}}\,(\rm{mag})$\tablenotemark{b} & 0.015 & 0.037 & 0.014       \\
	\enddata
	\tablecomments{All observations were conducted with the WFC3-IR instrument and F140W filter. $\lambda_{\rm{rest}}=\lambda_{\rm{obs}} / (1+z)$}
	\tablenotetext{a}{Determined from the GRB Afterglow.}
	\tablenotetext{b}{Foreground extinction calculated using \url{https://ned.ipac.caltech.edu/forms/calculator.html}} 
\end{deluxetable*}

\subsection{Photometry}
\label{sub:3.1}
\begin{figure}
	\epsscale{1.0}
	\plotone{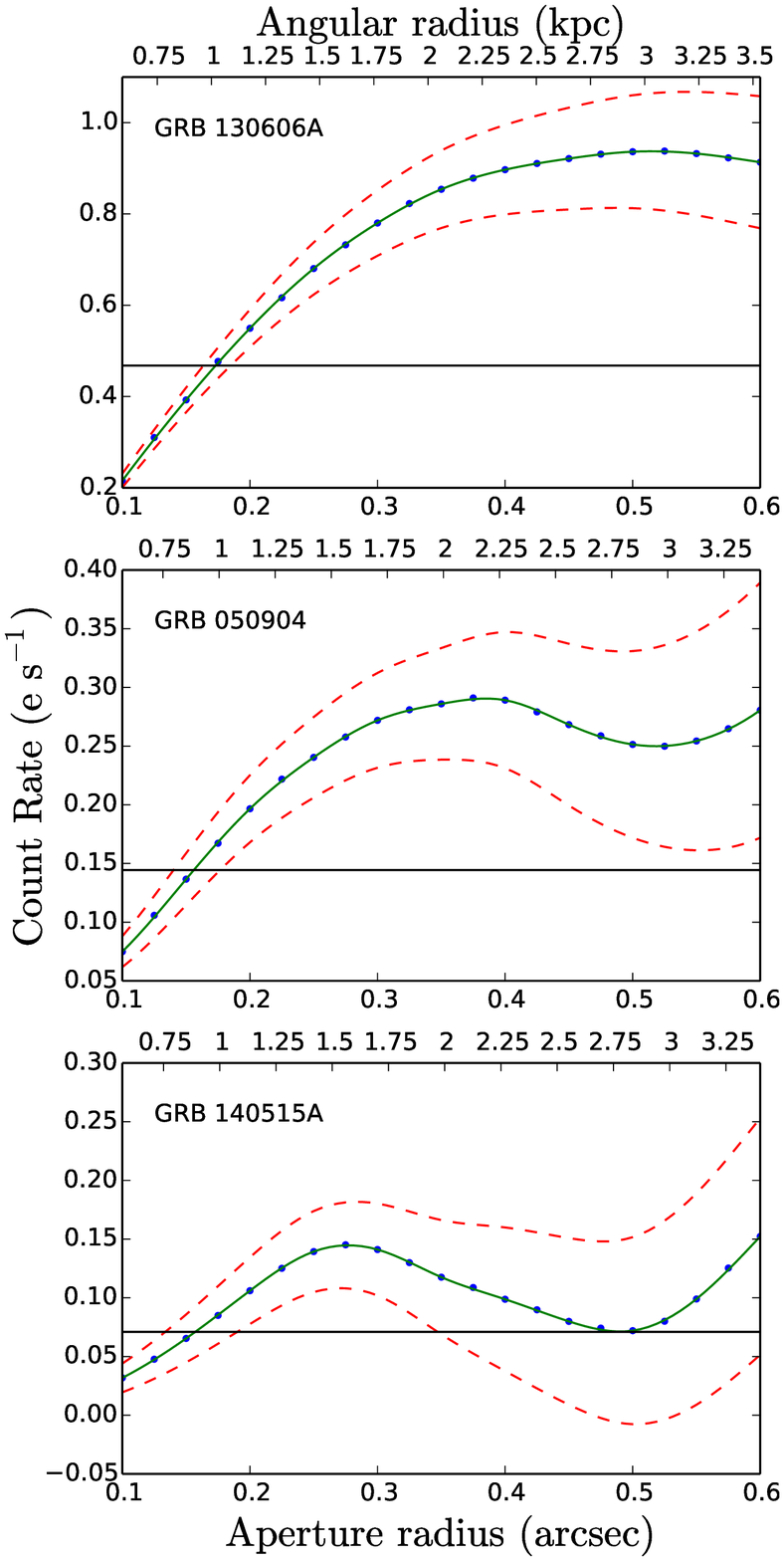}
	\caption{Photometry curves of growth for each host galaxy. Non-PSF and aperture-corrected values are shown as (blue) points, along with their (green) solid cubic spline interpolation curves. Photometric errors are shown as (red) dashed cubic spline interpolation curves. Half-maximum count rates are shown as (black) horizontal lines. Note that at larger radii the count rate is affected by light spilling in to the aperture from neighbouring sources.\label{fig:2}}
\end{figure}

Photometry of the three galaxies was performed using {\tt GAIA}\footnote{\url{http://star-www.dur.ac.uk/~pdraper/gaia/gaia.html}} and its {\tt Autophotom} package. A catalogue of objects from each {\em HST} image was built with {\tt GAIA's} native {\tt SExtractor} \citep{bear96} support, employing a detection threshold of $2.5\sigma\,\rm{pixel}^{-1}$ and requiring three neighbouring pixels above the background, to be considered objects. Identifying the objects located at each of the GRBs positions as the potential host galaxies, their count rates were measured using circular apertures centred on their barycentric catalogue coordinates. Sky background was estimated with $40$ circular apertures of identical radii to the galaxy aperture, distributed within $6\arcsec$ of the galaxy aperture, and taking care to avoid other objects detected in the catalogue to minimalise background contamination. Standard deviation of the background aperture count rates was adopted as the photometric error, which is reasonable, as we are in a background-limited regime. 

With a small sample, we chose to customize each source aperture radius instead of applying the more typical fixed aperture radius for all sources. Aperture radii were varied incrementally, with the background-subtracted count rates plotted into curves of growth (CoG) for each GRB host galaxy (Figure~\ref{fig:2}). Cubic spline interpolation was used to estimate values between the measured aperture radii. The total count rate can be estimated from the level at which the CoG plateaus (these peak radii are shown as the green circles in Figure~\ref{fig:1}). Although taking the peak of the CoG might systematically overestimate brightness, given that at larger radii we begin to be contaminated by neighbouring sources, our approach seems reasonable. As a compromise, we chose to average the three points at the peak of the CoG to represent the point at which the source counts reach the background level. Finally, we extrapolated the count rate using the tabulated encircled energy fraction\footnote{\url{http://www.stsci.edu/hst/wfc3/documents/handbooks/currentIHB/c07\_ir07.html}} and adopted the WFC3-IR infinite aperture zero point\footnote{\url{http://www.stsci.edu/hst/wfc3/phot\_zp\_lbn}} for the F140W filter. 

Based on Galactic foreground extinction maps from \citet{schl11}, we find for each case that only small corrections need to be applied (see Table~\ref{tb:1}). At the observed pivot wavelength, we estimate apparent magnitudes of $m_{\lambda_{\rm{obs}}}= 26.34^{+0.14}_{-0.16}, 27.56^{+0.18}_{-0.22},$ and $28.30^{+0.25}_{-0.33}$ respectively. For GRB 050904, the new result is consistent with the previous upper limits as discussed in Section~\ref{sub:2.2}. The most marginal detection is the case of GRB 140515A, but even here the significance is $\sim4\sigma$, giving us confidence that it is a real source. Subsequent photometric analysis and results are discussed in Section~\ref{sec:4} and given in Table~\ref{tb:2}.

\subsection{Size estimation}
\label{sub:3.2}
\citet{cul14} recently studied a sample of LBGs from {\em Hubble} deep imaging, in the $z=4$ to $z=8$ redshift range for their rest-frame UV sizes. They quantified galaxy size as the circularised half-light radius ($R_{\rm{half}}$), the radius enclosing half the galaxies' total flux (cf. Section~\ref{sub:3.1}). Following their methodology, we determine $R_{\rm{half}}$ from our CoGs, taking the interpolated aperture radii that intersect the estimated half-maximum count rates. \citet{cul14} used fixed apertures of radius $0\farcs6$. We accounted for each field's point spread function (PSF) by fitting a central Gaussian profile to bright, unsaturated, uncontaminated stars and averaging the results. 

We measured $\sigma_{\rm{PSF}}=0\farcs106$ for the field of GRB 130606A, and $\sigma_{\rm{PSF}}=0\farcs127$ for the fields of GRBs 050904 and 140515A. There is also an additional correction to account for the wings of the PSF which is estimated using the simulations of \citet{cul14}. After the total PSF correction, our half-light radii are determined to be $R_{\rm{half}}=0.15^{+0.02}_{-0.02}, 0.11^{+0.03}_{-0.03},$ and $0.12^{+0.05}_{-0.04}\,\rm{arcsec}$ for GRBs 130606A, 050904, and 140515A respectively. 

\subsection{Possible afterglow contamination?}
\label{sub:3.3}
Our observations of the field of GRB 140515A were obtained only $8.5\,\rm{months}$ months after the GRB event, and it is therefore possible, that our photometry could be contaminated by the fading afterglow. A reasonable upper limit to this contamination can be obtained by taking the flux in a small ($0\farcs1$ radius) aperture at the location of the afterglow, and assuming this is entirely due to a point source. This leads to a maximum contamination of $\approx4\,\rm{nJy}$. Alternatively, considering the latest reported infrared photometry, $m_{J,H} = 20.9$ \citep{mel15}, obtained at $\approx17\,\rm{hr}$ post-burst, and a typical late-time power-law decay of $F\propto\,t^{-1.5}$ \citep[see e.g., Tanvir et al. 2016 in prep.][]{kan10}, we would expect an afterglow flux density at the time of observation of $F\sim2\,\rm{nJy}$. We adopt this latter figure as the best compromise correction for afterglow contamination and apply it to our reported photometric results in Table~\ref{tb:2}, Section~\ref{sub:4.3} and Figures~\ref{fig:3}-\ref{fig:5}. We note that this correction is below the level of the photometric error, although it does reduce the overall significance of the detetcion to $\sim3.3\sigma$.

The other two fields were observed much longer post-GRB, and indeed the hosts in these cases are brighter, so any afterglow contamination should be negligible.

\subsection{Alternative photometric measurements}
\label{sub:3.4}
To verify that the photometric analysis presented here is robust and the results are independent of the specific procedure followed, we carried out an independent analysis of the images using the pipeline developed by the Brightest of Reionizing Galaxies (BoRG) survey, which searched for high-$z$ galaxies \citep[see][for a detailed description]{tre12,bra12,schm14}. In short, we first derived variance (RMS) maps from the weight maps produced by {\tt AstroDrizzle}, then ran {\tt SExtractor} in dual-image mode for a preliminary identification of the sources. Finally, we normalized the RMS maps to account for correlated noise by measuring the background in random positions that are not associated to any source, and ran {\tt SExtractor} with the normalized RMS maps to construct the final catalogue.

For GRB 050904 and GRB 130606A, the procedure yielded results consistent with the optimized measurements of Table~\ref{tb:2}, within the $1\sigma$ photometric uncertainty using standard {\tt SExtractor} parameters for searches of high-$z$ faint galaxies. For GRB 140515A, a standard run to identify the host galaxy failed, but setting a lower $S/N$ threshold for detection and aggressive deblending identified the host galaxy at $S/N=3.05$. Since the source is close to the detection limit of the images, and is located near an extended structure, it is not surprising that a general source detection algorithm is not recovering it as efficiently as the primary photometric approach adopted in this paper. Nevertheless, this analysis confirms that there is an excess of flux at the GRB position in all three cases. 

\section{Analysis and Discussion}
\label{sec:4}
\begin{deluxetable*}{lccc}
	\tablecolumns{4}
	\tablewidth{0pt}
	\tablecaption{Summary of host galaxy properties.\label{tb:2}}
	\tablehead{
	\colhead{Identifier} & \colhead{130606A} & \colhead{050904} & \colhead{140515A\tablenotemark{a}}
	} 
	\startdata
	$m_{\lambda_{\rm{obs}}}\,(\rm{mag})$                           & $26.34^{+0.14}_{-0.16}$  & $27.56^{+0.18}_{-0.22}$  & $28.30^{+0.25}_{-0.33}$ \\
	$M_{\lambda_{\rm{rest}}}\,(\rm{mag})$                          & $-20.38^{+0.14}_{-0.16}$ & $-19.26^{+0.18}_{-0.22}$ & $-18.36^{+0.29}_{-0.39}$\\
	$F_{\lambda_{\rm{obs}}}\,(\rm{nJy})$                           & $105\pm15$               & $34.3\pm6.3$             & $15.0\pm4.5$            \\
	$L_{1600} / L^{*}_{(z=6)}$                                     & $0.58$                   & $0.21$                   & $0.10$                  \\
	$R_{\rm{half}}\,(\rm{kpc})$                                    & $0.88^{+0.11}_{-0.09}$   & $0.64^{+0.19}_{-0.19}$   & $0.68^{+0.32}_{-0.29}$  \\
	$P_{\rm{cc}}\,(\%)$                                            & $1.4$                & $1.4$                & $2.3$               \\
	$\rm{Offset}$\tablenotemark{b}$\,(\arcsec)$                    & $0.06\pm0.02$            & $0.13\pm0.04$            & $0.21\pm0.07$           \\ 
	Galaxy RA\tablenotemark{c} ($^{\rm{h}}, ^{\rm{m}}, ^{\rm{s}}$) & 16:37:35.14              & 00:54:50.88              & 12:24:15.51             \\
	Galaxy Dec\tablenotemark{c} (\arcdeg, \arcmin, \arcsec)        & +29:47:46.5              & +14:05:09.9              & +15:06:16.8             \\	
	\enddata
	\tablecomments{$\lambda_{\rm{obs}}=13920\,\rm{\AA}$. $\lambda_{\rm{rest}}$ for each galaxy is given in Table~\ref{tb:1}.}
	\tablenotetext{a}{The values for 140515A are corrected for the possible afterglow contamination (see Section~\ref{sub:3.3}), with the exception of $m_{\lambda_{\rm{obs}}}$.}
	\tablenotetext{b}{Offset is defined as the projected straight-line distance between the GRB and galaxy centroid. Using our adopted cosmology these correspond to $0.35\pm0.12, 0.74\pm0.23$ and $1.19\pm0.40\,\rm{kpc}$ respectively.}
	\tablenotetext{c}{J2000 coordinates of the detected host galaxies.}
\end{deluxetable*}

\subsection{Are they the host galaxies?}
\label{sub:4.1}
It is possible that the three detected objects are not the host galaxies of each GRB, but are instead intervening objects at a lower redshift. Deep imaging blueward of the Ly$\alpha$ break is currently not available, and we cannot confirm their drop-out nature. Despite this, we can estimate the probability of chance coincidence ($P_{\rm{cc}}$) of a lower redshift galaxy, based on galaxy number counts \citep[as has been used for both long- and short-GRB hosts in the past e.g.,][]{fon13}. Following the method of \citet{blo02}:

\begin{equation}
	P_{\rm{cc}} = 1 - \exp[-4\pi R_{\rm{half}}^{2}\sigma(\ge m)],
	\label{eq:1}
\end{equation}
\noindent where $\sigma(\ge m)$ is the observed number density of galaxies brighter than magnitude $m$. This is the appropriate formulation when a GRB afterglow is localized within the detectable light of its host galaxy, as is the case here (Figure~\ref{fig:1}). The projected offsets from the host centres are shown in Table~\ref{tb:2}, and this small sample is consistent with the roughly $50\%$ fraction of bursts found within the (blue light) half-light radius by \citet{blo02}. 

Interpolating $H$-band number counts measured by \citet{met06} and using our calculated $R_{\rm{half}}$ measurements, we determine $P_{\rm{cc}}$ values of $\la2\,\%$ (see Table~\ref{tb:2}) for all three objects. This statistic indicates that these are all likely the GRB host galaxies, and that the probability that two or even all three detections are lower redshift interlopers is very low.

An alternative approach to addressing this question is to directly measure the fraction of our images covered by visible galaxies. To do this we ran SExtractor to find all non-stars (using the same parameters discussed in Section~\ref{sub:3.1}). Summing their total area, we find that only $\approx0.5$\% of each {\em HST} image coincides with detectable galaxies, suggesting that the $P_{\rm cc}$ values evaluated above are, if anything, rather conservative for our fields.

As is common in spectroscopy of high redshift continuum sources, absorption lines of intervening systems at lower redshifts are seen in all three GRB afterglows studied here (see Section~\ref{sec:2}). These absorption systems (in particular \ion{C}{4} at $z\ga2$), are typically associated with galaxies with impact parameters that are tens of kpc (many arcseconds) from the line of sight \citep{ade05}. Thus, {\em a priori}, it would be surprising if the galaxy responsible for absorption happened to coincide spatially with the GRB location. Indeed there is no evidence for particularly strong absorption or dust attenuation that would be suggestive of these GRBs being directly behind a lower redshift galaxy. 

\subsection{Comparison with Lyman-break Galaxies}
\label{sub:4.2}
\begin{figure}
	\epsscale{1.2}
	\plotone{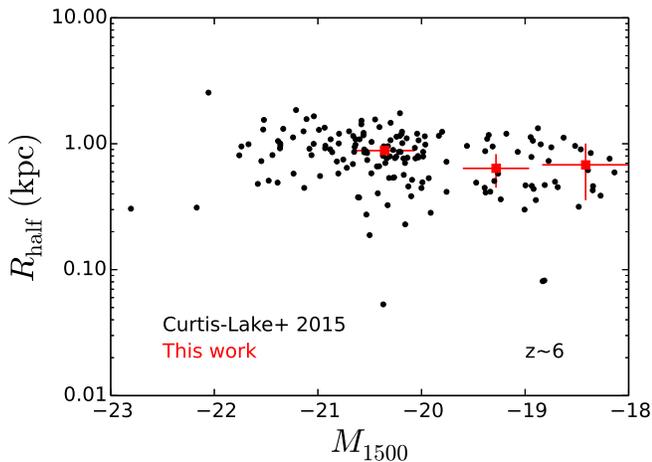}
	\caption{Plot of half-light radius vs. absolute magnitude for our host galaxies (red squares), compared to the sample of \citet{cul14} LBGs at $z\sim6$.\label{fig:3}}
\end{figure}

Converting our apparent magnitudes from the observed pivot wavelength to literature comparison wavelengths of $1500-1600\,\rm{\AA}$ requires knowledge of the underlying spectrum and its spectral index ($\beta_{\rm{UV}}$), where the UV continuum is typically assumed to be a power-law with $f_{\lambda}\propto\lambda^{\beta_{\rm{UV}}}$. \citet{dun15} compare a variety of UV continuum studies and find little difference between $z\sim6$ and $z\sim7$ (see their Figure 2) in the parameterization of the UV spectral index, 

\begin{equation}
	\beta_{\rm{UV}} = (-2.05\pm0.04) + (-0.13\pm0.04)\times(M_{\lambda_{\rm{rest}}} + 19.5).
	\label{eq:2}
\end{equation}

We assume that this parametrization is applicable to our host galaxies and find $\beta_{\rm{UV}}$ values in the range $-2.2$ to $-1.9$. We also adopt an intrinsic scatter of $\sigma_{\beta_{\rm{UV}}}\approx0.25$, based on the results of \citet{rog14}. Magnitudes are converted using the following equation:

\begin{equation}
	M_{\lambda} = M_{\lambda_{\rm{obs}}} - 2.5 \times (\beta_{\rm{UV}} + 2)\times \log((1+z) \times \lambda/\lambda_{\rm{obs}}).
	\label{eq:3}
\end{equation}

We estimate absolute magnitudes ($M_{1600}$) in the range $-20.5$ to $-18.5$, a difference of only $0.02$ to $0.03$ from the observed wavelength (absolute) magnitudes. Considering the size of our observational errors, this difference is negligible. 

\citet{bou15} present results for the UV galaxy luminosity function (LF) from $z=4$ to $z=10$, based on large samples of LBGs with photometric redshifts. The LF is conventionally described using the Schechter function \citep{sche76}:

\begin{equation}
	LF(x)dx = \phi^{*}x^{-\alpha}e^{-x}dx,
	\label{eq:4}
\end{equation}
		
\noindent where $x = L/L^{*}$, with $L^{*}$ being the characteristic luminosity break between the faint-end power-law slope ($\alpha$) and the bright-end exponential cut-off, while $\phi^{*}$ is a normalization factor. \citet{bou15} present a parameterization of $M_{1600}^{*}$ (the magnitude corresponding to $L^{*}$) using a linear fit as a function of redshift:

\begin{equation}
	M_{1600}^{*}=(-20.95\pm0.10)+(0.01\pm0.06)\times(z-6).
	\label{eq:5}
\end{equation}

From $z=5.913$ to $z=6.327$, $M_{1600}^{*}$ varies only by $0.004$ mag, and hence we adopt the value $M_{1600}^{*}=-20.95\pm0.12$. Converting our GRB host galaxies into comparable ratios gives $0.1-0.6\,L^{*}_{z=6}$ (see Table~\ref{tb:2}). This range is consistent with that found in deep LBG samples given the low extinction inferred from the afterglows, these galaxies would likely also satisfy the colour selection criteria for inclusion in these LBG samples \citep{bou15}. 

Using our adopted cosmology, we convert our $R_{\rm{half}}$ values into kpc, giving the range $0.6-0.9\,\rm{kpc}$ (see Table~\ref{tb:2}). We plot these sizes against magnitudes in Figure~\ref{fig:3}, together with the LBG sample of \citet{cul14}. All three GRB host galaxies have typical $R_{\rm{half}}$ for their magnitudes. We also note that our GRB host galaxies have comparable $R_{\rm{half}}$ values to those of Ly$\alpha$ selected galaxies at $z\sim6$ \citep{mal12}.

\subsection{Star Formation Histories}
\label{sub:4.3}
\begin{figure}
	\epsscale{1.2}
	\plotone{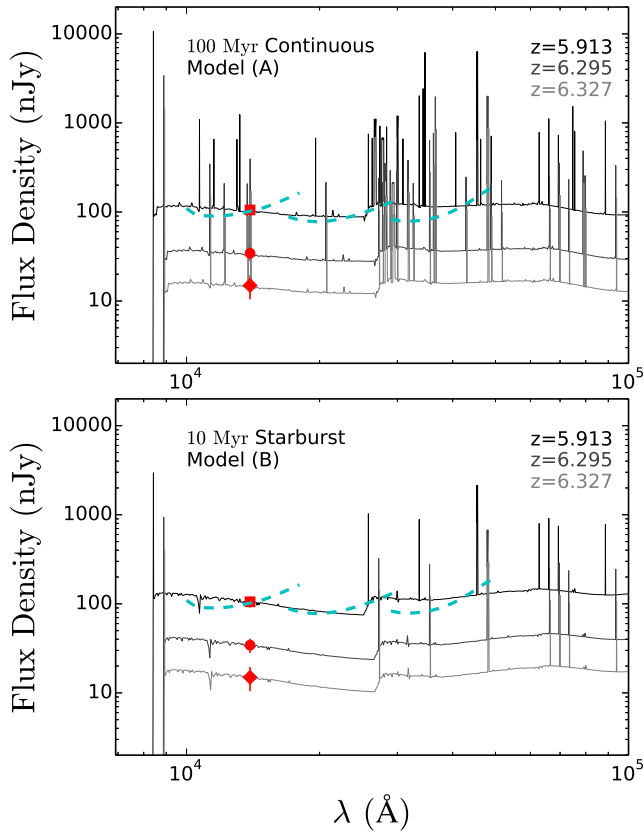}
	\caption{BPASS SED models redshifted and scaled to the each GRB host luminosity (see Section~\ref{sub:4.3}). Our $F_{\lambda_{\rm{obs}}}$ for GRB 130606A (square), GRB 050904 (circle), and GRB 140515A (diamond) are plotted as reference. {\em JWST's} NIRSpec sensitivity (dashed cyan) for resolution, $R=1000$, spectroscopy is also shown. NIRSpec is modelled to an exposure of $10^{4}\,\rm{s}$, and at a signal-to-noise ratio of $3$.\label{fig:4}}
\end{figure}

The host galaxies were observed in the rest-frame $1900-2000\,\rm{\AA}$ range, where the emission is dominated by the light of young, massive stars. Therefore, our photometry provides constraints on their star formation histories. To investigate this, we employ the BPASS stellar population synthesis models \citep{eld09,sta15}, which incorporate prescriptions for binary stellar evolution. We also choose to adopt their preferred broken power-law model for the differential stellar initial mass function (IMF), which has a slope of $-1.3$ between $0.1-0.5\,\rm{M}_{\odot}$ and a slope of $-2.35$ between $0.5-100\,\rm{M}_{\odot}$. Given the abundance constraints discussed in Section~\ref{sec:2}, from the GRB afterglow spectra, we use the lowest metallicity ($Z=0.001$; i.e. $\sim5\%\,Z_{\odot}$) BPASS models.

We consider two cases which span a broad range of plausible histories. Model (A) has a continuous star formation rate for $100\,\rm{Myr}$, which is reasonable given the age of the universe of $\sim900\,\rm{Myr}$ at $z\sim6$. Our alternative model (B) assumes just a single short-lived burst of star formation $10\,\rm{Myr}$ prior to the observation epoch, which would be reasonable if the GRB progenitors were born in the same starburst. Their likely high masses we believe they had, would imply a lifetime of this order. It is interesting to note that despite its young age, this model results in significant near-IR emission due to the contribution of red supergiants.

Starting with the spectral energy density (SED) BPASS models\footnote{\url{http://www.bpass.org.uk/}}, we folded these through the radiative transfer program {\tt Cloudy}\footnote{\url{http://www.nublado.org}} \citep{fer98}, to obtain the underlying nebular continuum and line emission spectrum excited by the BPASS stellar spectra. Next, we scaled the emission lines to an intrinsic spectral resolution of R = 1000. Finally, we transformed these models to the observed redshifts of the bursts, and scaled them to our photometric results. For model (A), our inferred star formation rates, are $SFR=5.3\pm0.7, 1.8\pm0.3,$ and $0.8\pm0.2\,\rm{M_{\odot}\,yr^{-1}}$, while the total mass of stars formed instantaneously in model (B) are $\mathcal{M}=24.3\pm3.4, 8.4\pm1.5,$ and $3.7\pm1.1\,\times10^7\,\rm{M_{\odot}}$ for the hosts of GRBs 130606A, 050904, and 140515A respectively. These scaled models are shown for each GRB host in Figure~\ref{fig:4}.

The above estimates implicitly assume negligible dust extinction, as is frequently done for galaxies at these redshifts \citep[although see][who argue for modest UV extinctions of $0.35-0.5\,\rm{mag}$ based on colours of $z\sim7$ LBGs]{wil13}. For our cases, the observations of the afterglows support the proposition that the dust content is generally low, with $A_{\rm{V}}\la0.1\,\rm{mag}$ in each case (see Section~\ref{sec:2}). If we take these afterglow dust estimates as representative of the internal extinction in their hosts, we can find rest-frame extinctions at $\sim2000\,\rm{\AA}$, as observed. Depending on the adopted extinction law, we find $A_{2000}\approx2-3\,A_{\rm{V}}$ \citep{pei92,cal20}. Hence our SFR values could increase by up to $\sim30\%$ if corrected for dust extinction.  

\subsection{Comparison to other GRB hosts}
\label{sub:4.4}
\begin{figure}
	\epsscale{1.2}
	\plotone{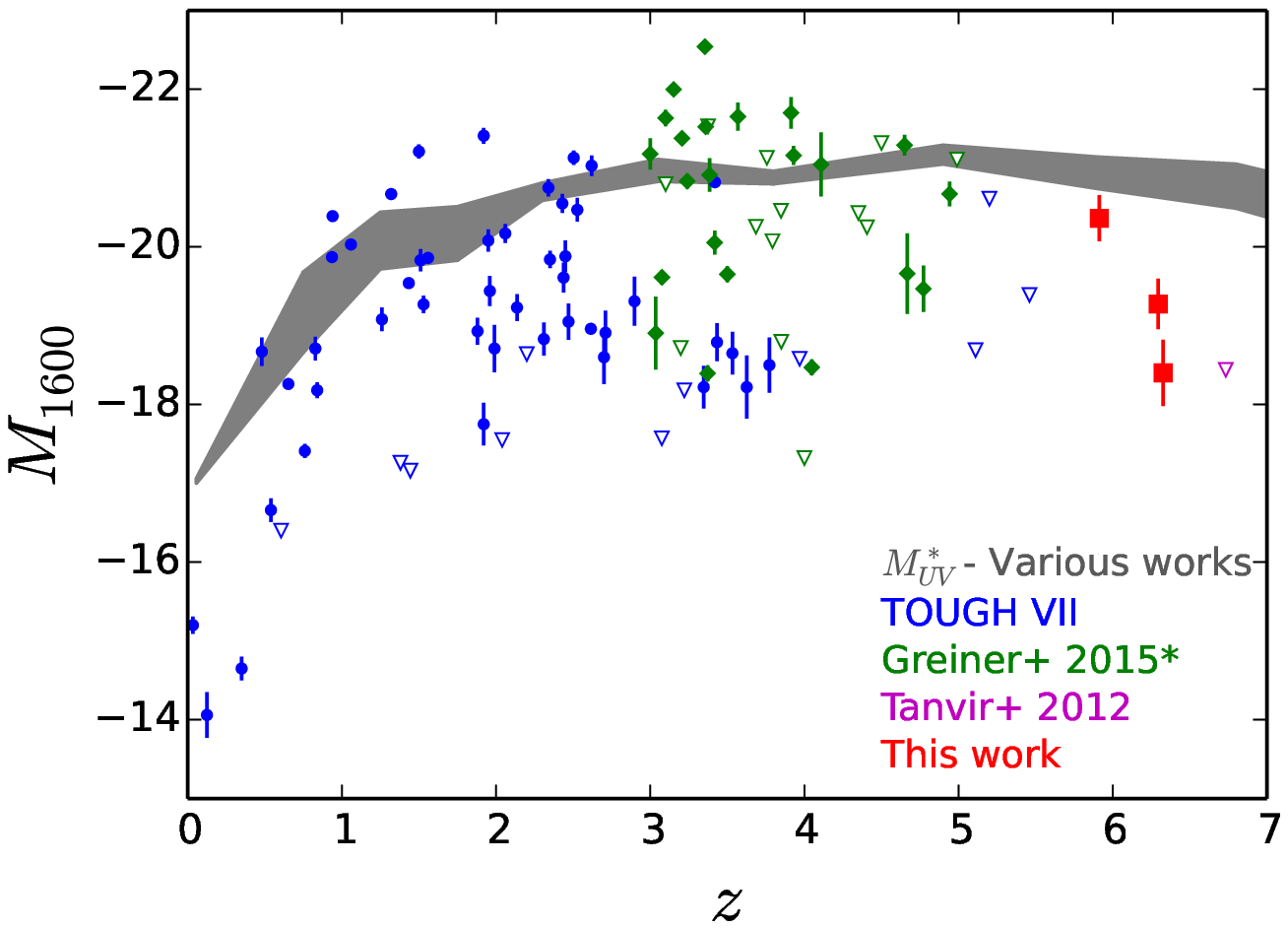}
	\caption{Our host galaxies (red squares) and their $1\sigma$ limits, plotted alongside the TOUGH (blue points) \citet{schu15}, \citet{gre15} (green diamonds; recalculated as described in Section~\ref{sub:4.4}) and \citet{tan12} (magenta triangles) samples of GRB hosts, showing evolution of the population with redshift. Non-detections (triangles) are given at $3\sigma$. Schechter function $M_{1600}^{*}$ values representing the UV LF are shown as the (grey) shaded region determined from \citet{red09,oes10,rob11} and \citet{bou15}. Uncertainty ($1\sigma$) in $M_{1600}^{*}$ is represented by width of the (grey) region. \label{fig:5}} 
\end{figure}

It is interesting to compare the properties of our detected $z\sim6$ hosts with the population of GRB hosts at lower redshifts. One problem in doing this is that samples of GRBs and their hosts tend to be inhomogeneous, incomplete, and biased, in particular against GRBs occurring in dusty galaxies for which no optical afterglows were found \citep[e.g.,][]{per13}. The Optically Unbiased GRB Host (TOUGH) survey attempted to define a more statistically representative sample of hosts by identifying them using their X-ray localisations, and hence obtaining redshifts directly from the hosts, where an afterglow redshift was unavailable. The net result is a sample of $69$ hosts for which $61$ have redshifts and all but $1$ of the remainder have photometric constraints placing them at $z\la6$ \citep{hjo12,jak12}. 

Beyond $z\sim3$ the TOUGH sample suffers from small number statistics, and we supplement it with additional hosts for which photometry is available in the literature. Although these additional data are more subject to optical selection effects, and include some non-{\em Swift} GRBs, the larger sample size likely does provide a fuller picture of the host luminosity distribution at higher redshifts. The $1600\,\rm{\AA}$ UV luminosities of TOUGH VII \citep{schu15}, \citet{gre15} and \citet{tan12} are plotted in Figure~\ref{fig:5} together with our new hosts. To ensure that the samples of TOUGH VII and \citet{gre15} are comparable, we re-analyzed the latter sample using the methodology of TOUGH VII \citep{schu15}. We also plot $M_{1600}^{*}$ as a function of redshift. At low redshifts ($z\la1$) the bulk of hosts are sub-$L^{*}$, as expected if GRBs trace primarily sub-solar metallicity star formation \citep[e.g.,][]{lev10,gra13,tre15,ver15}. From $z\sim1$ to $z\sim6$ the break of the luminosity function evolves rather little. While some hosts are found with $L>L^{*}$ up to $z\sim4$, at higher redshifts such bright hosts are absent, consistent with a steepening galaxy luminosity function \citep[c.f.,][]{bou15}, with star formation predominantly occurring in faint galaxies.

\section{Conclusions}
\label{sec:5}
We have presented deep ($10-13\,\rm{ks}$) {\em HST}/F140W observations of the fields of GRBs 130606A, 050904, and 140515A and have identified galaxies coincident with each GRB afterglow location. Based on low ($\la2\,\%$) chance of coincidence of low-$z$ interlopers, we conclude that these are very likely the GRB host galaxies. At $z\sim6$, these are the most distant GRB host galaxies detected to date. It is significant that we are now detecting GRB hosts at the end of the era of reionization (a fuller statistical analysis of the high-z GRB hosts, including non-detections, will be presented in a future paper, in which we will address their implications for the galaxy luminosity function at $z>6$).

With rest-frame UV ($\lambda\sim2000\,\rm{\AA}$) luminosities in the range of $0.1-0.6\,L^{*}_{z=6}$, and half-light radii of $0.6-0.9\,\rm{kpc}$, our three host GRB galaxies are consistent with those found from deep {\em HST} studies of LBGs in the same redshift range. We find that the GRBs themselves are located within the rest-frame UV light of their hosts, as is usually the case for GRBs at lower redshifts \citep{fru06,sve10}. 

The power of GRB selection is that spectroscopy of the afterglows can provide detailed information about the gas and dust properties of the hosts. In this instance, it tells us that all three galaxies have low metallicity ($[\rm{M/H}]\la-1$) and low dust extinction ($A_{\rm{V}}\la0.1$). These galaxies are considerably brighter than the upper limits for the five $z>5$ hosts (excluding GRB 050904 itself) reported by \citet{tan12}. Of course, these studies remain limited to relatively small samples, but it would  have been surprising had our new program not begun to make some detections, given that canonical LFs suggest that $>20\%$ of star formation at $z\sim6$ should be occurring in galaxies brighter than $\approx20\,\rm{nJy}$ \citep{tan12}. The small sample size also precludes assessing whether there are any correlations between the properties of the hosts in emission and their properties derived from the afterglows, although it is interesting to note that GRB 140515A, whose afterglow revealed a low foreground gas column, is also the faintest of the three hosts and the one where the burst occurred at the greatest offset from the host centre.

Significantly, as illustrated in Figure~\ref{fig:4}, we would expect {\em James Webb Space Telescope} spectroscopy to provide high signal-to-noise measurements of the rest-frame optical emission lines for potentially all three hosts, especially GRB 130606A. This will allow comparison of emission-line metallicity and extinction diagnostics with the detailed values obtained from the afterglows, which are potentially important for testing the applicability of these diagnostics to this redshift regime. 

\acknowledgments
We acknowledge support from the UK Science and Technology Facilities Council (STFC) and the HST-GO-13831.002-A grant from STScI. The research leading to these results has received funding from the European Research Council (ERC) under the European Union's Seventh Framework Pogram (FP7/2007-2013)/ERC Grant agreement no. EGGS-278202. Based on observations made with the NASA/ESA {\em Hubble Space Telescope}. STScI is operated by the association of Universities for Research in Astronomy, Inc. under the NASA contract NAS5-26555. These observations are associated with {\em HST} program GO-13831 (PI:Tanvir). SSchulze acknowledges support from CONICYT-Chile FONDECYT 3140534, Basal-CATA PFB-06/2007, and Project IC120009 "Millennium Institute of Astrophysics (MAS)" of Inciativa Cient\'{\i}fica Milenio del Ministerio de Econom\'{\i}a, Formento y Turismo. EC-L acknowledges financial support from the ERC via an Advanced Grant under grant agreement no. 321323-NEOGAL.

{\it Facility:} \facility{HST (WFC3)}

\appendix
\section{Relative Astrometry}
\label{apsec:1}
For our analysis it is important to establish the locations at which the GRBs occurred on our {\em HST}/WFC3 images, at least well enough to give confidence in the identification of the hosts. To achieve this we made use of observations made shortly after the GRB events, on which the afterglows were detected, and calculated the transformation between these images and the {\em HST} frames using a number of field sources. The precision with which this can be achieved depends on a number of factors, in particular the detection significance of the afterglows and the number and brightness of field sources visible on the {\em HST} frames and comparison frames. The precision with which the afterglows can be centroided is then estimated by:

\begin{equation}
	\sigma_{\rm AG}\approx\sigma_{\rm PSF}/SNR 
	\label{apeq:1}
\end{equation}
\noindent where $SNR$ is the photometric detection significance (here obtained from aperture photometry). The error introduced by the coordinate transformation is best estimated by the object to object scatter (the plate scale, orientation and distortion mapping should be reliable from the pipeline processing of each image, especially over the small {\em HST} fields, and we therefore only fit for a zero-point translation). Here we summarise the comparison images and tabulate the error budget in each case (Table~\ref{aptb:1}).

\begin{deluxetable*}{lccccc}
	\tablecolumns{6}
	\tablewidth{0pt}
	\tablecaption{Ground-based afterglow astrometry error budget.\label{aptb:1}}
	\tablehead{
	\colhead{} & \multicolumn{3}{c}{GRB 130606A} & \colhead{GRB 050904} & \colhead{GRB 140515A} \\
	\colhead{Identifier} & \colhead{$Z$-band} & \colhead{$J$-band} & \colhead{$K$-band} &
	\colhead{$JHK$-band} & \colhead{$z$-band} 
	}
	\startdata
	Std error in matching coordinate systems$\,(\arcsec)$ & 0.014 & 0.014 & 0.016 & 0.031 & 0.018 \\
	Afterglow centroiding error$\,(\arcsec)$ & 0.012 & 0.009 & 0.007 & 0.017 & 0.002 \\
	Combined error$\,(\arcsec)$              & 0.018 & 0.016 & 0.018 & 0.035 & 0.018 \\
	\enddata
	\tablecomments{All three image bands for GRB 050904 were co-added to increase signal-to-noise.}
\end{deluxetable*}

\subsection{GRB 130606A}
\label{apsub:1}
Our primary comparison images were obtained with the United Kingdom Infrared Telescope (UKIRT) Wide Field Camera (WFCAM) in the $Z$, $J$ and $K$ bands. These wide field images are astrometrically calibrated to the Two Micron All-Sky Survey (2MASS) system. We made use of $12$ field stars which were well detected in the WFCAM images and visible in the {\em HST} fields.

In fact the independent estimates of the afterglow position from the three bands agree to better than this error, so we average them to obtain the final best estimate.

\subsection{GRB 050904}
\label{apsub:2}
In this case our primary astrometric comparison was with several UKIRT/WFCAM images taken in the $J$, $H$ and $K$ bands, which we co-added to increase the signal-to-noise. A rather higher scatter in the mapping to the {\em HST} image was noticed for stars in this field, possibly due to small proper motions during the intervening decade from the burst occurrence. We therefore utilized $4$ compact galaxies as well as $3$ stars to make the astrometric comparison.

We also analyzed the {\em HST}/NICMOS image reported by \citet{ber07}, finding a consistent location within the errors, although the signal-to-noise of the afterglow was rather low.

\subsection{GRB 140515A}
\label{apsub:3}
We used Gemini-North Gemini Multi-Object Spectrograph (GMOS) imaging as the astrometric comparison, utilising $8$ stars that were also on our {\em HST} images. The seeing was good on the GMOS image ($\approx0\farcs5$), and the signal-to-noise ratio high ($\approx100$).


\begin{thebibliography}{}
\bibitem[Adelberger et al.(2005)]{ade05} Adelberger, K.~L., Shapley, A.~E., Steidel, C.~C., et al.\ 2005, \apj, 629, 636 
\bibitem[Basa et al.(2012)]{bas12} Basa, S., Cuby, J.~G., Savaglio, S., et al.\ 2012, \aap, 542, A103
\bibitem[Berger et al.(2007)]{ber07} Berger, E., Chary, R., Cowie, L.~L., et al.\ 2007, \apj, 665, 102
\bibitem[Bertin \& Arnouts(1996)]{bear96} Bertin, E., \& Arnouts, S.\ 1996, \aaps, 117, 393 
\bibitem[Bloom et al.(2002)]{blo02} Bloom, J.~S., Kulkarni, S.~R., \& Djorgovski, S.~G.\ 2002, \aj, 123, 1111 
\bibitem[Bloom et al.(2009)]{blo09} Bloom, J.~S., Perley, D.~A., Li, W., et al.\ 2009, \apj, 691, 723  
\bibitem[Bouwens et al.(2010)]{bou10} Bouwens, R.~J., Illingworth, G.~D., Oesch, P.~A., et al.\ 2010, \apjl, 708, L69 
\bibitem[Bouwens et al.(2015)]{bou15} Bouwens, R.~J., Illingworth, G.~D., Oesch, P.~A., et al.\ 2015, \apj, 803, 34 
\bibitem[Bradley et al.(2012)]{bra12} Bradley, L.~D., Trenti, M., Oesch, P.~A., et al.\ 2012, \apj, 760, 108 
\bibitem[Calzetti et al.(2000)]{cal20} Calzetti, D., Armus,  L., Bohlin, R.~C., et al.\ 2000, \apj, 533, 682 
\bibitem[Campana et al.(2015)]{cam15} Campana, S., Salvaterra, R., Ferrara, A., \& Pallottini, A.\ 2015, \aap, 575, A43 
\bibitem[Capak et al.(2015)]{cap15} Capak, P.~L., Carilli, C., Jones, G., et al.\ 2015, \nat, 522, 455
\bibitem[Castro-Tirado et al.(2013)]{cas13} Castro-Tirado, A.~J., S{\'a}nchez-Ram{\'{\i}}rez, R., Ellison, S.~L., et al.\ 2013, \aap, submitted (arXiv:1312.5631) 
\bibitem[Chary et al.(2007)]{cha07} Chary, R., Berger, E., \& Cowie, L.\ 2007, \apj, 671, 272 
\bibitem[Chornock et al.(2013)]{cho13} Chornock, R., Berger, E., Fox, D.~B., et al.\ 2013, \apj, 774, 26 
\bibitem[Chornock et al.(2014)]{cho14} Chornock, R., Berger, E., Fox, D.~B., et al.\ 2014, \apjl, submitted (arXiv:1405.7400) 
\bibitem[Cucchiara et al.(2011)]{cuc11} Cucchiara, A., Levan, A.~J., Fox, D.~B., et al.\ 2011, \apj, 736, 7 
\bibitem[Cucchiara et al.(2015)]{cuc15} Cucchiara, A., Fumagalli, M., Rafelski, M., et al.\ 2015, \apj, 804, 51 
\bibitem[Curtis-Lake et al.(2016)]{cul14} Curtis-Lake, E., McLure, R.~J., Dunlop, J.~S., et al.\ 2016, \mnras, 457, 440 
\bibitem[de Ugarte Postigo et al.(2012)]{dup12} de Ugarte Postigo, A., Fynbo, J.~P.~U., Th{\"o}ne, C.~C., et al.\ 2012, \aap, 548, A11 
\bibitem[Duncan \& Conselice(2015)]{dun15} Duncan, K., \& Conselice, C.~J.\ 2015, \mnras, 451, 2030 
\bibitem[Dunlop et al.(2012)]{dun12} Dunlop, J.~S., McLure, R.~J., Robertson, B.~E., et al.\ 2012, \mnras, 420, 901 
\bibitem[Eldridge \& Stanway(2009)]{eld09} Eldridge, J.~J., \& Stanway, E.~R.\ 2009, \mnras, 400, 1019
\bibitem[Ferland et al.(1998)]{fer98} Ferland, G.~J., Korista, K.~T., Verner, D.~A., et al.\ 1998, \pasp, 110, 761  
\bibitem[Finkelstein et al.(2012)]{fin12} Finkelstein, S.~L., Papovich, C., Ryan, R.~E., et al.\ 2012, \apj, 758, 93 
\bibitem[Fong et al.(2013)]{fon13} Fong, W., Berger, E., Chornock, R., et al.\ 2013, \apj, 769, 56 
\bibitem[Fruchter et al.(2006)]{fru06} Fruchter, A.~S., Levan, A.~J., Strolger, L., et al.\ 2006, \nat, 441, 463 
\bibitem[Fynbo et al.(2008)]{fyn08} Fynbo, J.~P.~U., Prochaska, J.~X., Sommer-Larsen, J., Dessauges-Zavadsky, M., \& M{\o}ller, P.\ 2008, \apj, 683, 321 
\bibitem[Fynbo et al.(2009)]{fyn09} Fynbo, J.~P.~U., Jakobsson, P., Prochaska, J.~X., et al.\ 2009, \apjs, 185, 526 
\bibitem[Gonz{\'a}lez et al.(2012)]{gon12} Gonz{\'a}lez, V., Bouwens, R.~J., Labb{\'e}, I., et al.\ 2012, \apj, 755, 148 
\bibitem[Gonz{\'a}lez et al.(2014)]{gon14} Gonz{\'a}lez, V., Bouwens, R., Illingworth, G., et al.\ 2014, \apj, 781, 34 
\bibitem[Graham \& Fruchter(2013)]{gra13} Graham, J.~F., \& Fruchter, A.~S.\ 2013, \apj, 774, 119 
\bibitem[Graham \& Fruchter(2015)]{gra15a} Graham, J.~F., \& Fruchter, A.~S.\ 2015, \apj, submitted (arXiv:1511.01079) 
\bibitem[Graham et al.(2015)]{gra15b} Graham, J.~F., Fruchter, A.~S., Levesque, E.~M., et al.\ 2015, \apj, submitted (arXiv:1511.00667) 
\bibitem[Greiner et al.(2015)]{gre15} Greiner, J., Fox, D.~B., Schady, P., et al.\ 2015, \apj, 809, 76 
\bibitem[Hartoog et al.(2015)]{har15} Hartoog, O.~E., Malesani, D., Fynbo, J.~P.~U., et al.\ 2015, \aap, 580, A139
\bibitem[Hjorth et al.(2012)]{hjo12} Hjorth, J., Malesani, D., Jakobsson, P., et al.\ 2012, \apj, 756, 187
\bibitem[Jakobsson et al.(2004)]{jak04} Jakobsson, P., Hjorth, J., Fynbo, J.~P.~U., et al.\ 2004, \aap, 427, 785  
\bibitem[Jakobsson et al.(2006)]{jak06} Jakobsson, P., Levan, A., Fynbo, J.~P.~U., et al.\ 2006, \aap, 447, 897 
\bibitem[Jakobsson et al.(2012)]{jak12} Jakobsson, P., Hjorth, J., Malesani, D., et al.\ 2012, \apj, 752, 62 
\bibitem[Kann et al.(2010)]{kan10} Kann, D.~A., Klose, S., Zhang, B., et al.\ 2010, \apj, 720, 1513 
\bibitem[Kawai et al.(2006)]{kaw06} Kawai, N., Kosugi, G., Aoki, K., et al.\ 2006, \nat, 440, 184  
\bibitem[Kr{\"u}hler et al.(2015)]{kru15} Kr{\"u}hler, T., Malesani, D., Fynbo, J.~P.~U., et al.\ 2015, \aap, 581, A125 
\bibitem[Laskar et al.(2013)]{las13} Laskar, T., Zauderer, A., \& Berger, E.\ 2013, GRB Coordinates Network, 14817, 1 
\bibitem[Laskar et al.(2014)]{las14} Laskar, T., Zauderer, A., \& Berger, E.\ 2014, GRB Coordinates Network, 16283, 1 
\bibitem[Levesque et al.(2010)]{lev10} Levesque, E.~M., Kewley, L.~J., Berger, E., \& Zahid, H.~J.\ 2010, \aj, 140, 1557
\bibitem[Malhotra et al.(2012)]{mal12} Malhotra, S., Rhoads, J.~E., Finkelstein, S.~L., et al.\ 2012, \apjl, 750, L36  
\bibitem[Melandri et al.(2015)]{mel15} Melandri, A., Bernardini, M.~G., D'Avanzo, P., et al.\ 2015, \aap, 581, A86 
\bibitem[Metcalfe et al.(2006)]{met06} Metcalfe, N., Shanks, T., Weilbacher, P.~M., et al.\ 2006, \mnras, 370, 1257
\bibitem[Oesch et al.(2010)]{oes10} Oesch, P.~A., Bouwens, R.~J., Carollo, C.~M., et al.\ 2010, \apjl, 725, L150 
\bibitem[Oesch et al.(2014)]{oes14} Oesch, P.~A., Bouwens, R.~J., Illingworth, G.~D., et al.\ 2014, \apj, 786, 108 
\bibitem[Oesch et al.(2015a)]{oes15a} Oesch, P.~A., Bouwens, R.~J., Illingworth, G.~D., et al.\ 2015, \apj, 808, 104 
\bibitem[Oesch et al.(2015b)]{oes15b} Oesch, P.~A., van Dokkum, P.~G., Illingworth, G.~D., et al.\ 2015, \apjl, 804, L30 
\bibitem[Oke \& Gunn(1983)]{oke83} Oke, J.~B., \& Gunn, J.~E.\ 1983, \apj, 266, 713
\bibitem[Pei(1992)]{pei92} Pei, Y.~C.\ 1992, \apj, 395, 130 
\bibitem[Perley et al.(2013)]{per13} Perley, D. A. et al., 2013, \apj, 778, 128
\bibitem[Perley et al.(2016)]{per15} Perley, D.~A., Tanvir, N.~R., Hjorth, J., et al.\ 2016, \apj, 817, 8 
\bibitem[Pian et al.(2006)]{pia06} Pian, E., Mazzali, P.~A., Masetti, N., et al.\ 2006, \nat, 442, 1011 
\bibitem[Pier et al.(2003)]{pie03} Pier, J.~R., Munn, J.~A., Hindsley, R.~B., et al.\ 2003, \aj, 125, 1559
\bibitem[Planck Collaboration(2015)]{pck15} Planck Collaboration, Ade, P.~A.~R., Aghanim, N., et al.\ 2015, \aap, submitted (arXiv:1502.01589) 
\bibitem[Racusin et al.(2008)]{rac08} Racusin, J.~L., Karpov, S.~V., Sokolowski, M., et al.\ 2008, \nat, 455, 183 
\bibitem[Reddy \& Steidel(2009)]{red09} Reddy, N.~A., \& Steidel, C.~C.\ 2009, \apj, 692, 778 
\bibitem[Robotham \& Driver(2011)]{rob11} Robotham, A.~S.~G., \& Driver, S.~P.\ 2011, \mnras, 413, 2570
\bibitem[Rogers et al.(2014)]{rog14} Rogers, A.~B., McLure, R.~J., Dunlop, J.~S., et al.\ 2014, \mnras, 440, 3714 
\bibitem[Salvaterra et al.(2009)]{sal09} Salvaterra, R., Della Valle, M., Campana, S., et al.\ 2009, \nat, 461, 1258 
\bibitem[Schady et al.(2012)]{scha12} Schady, P., Dwelly, T., Page, M.~J., et al.\ 2012, \aap, 537, A15 
\bibitem[Schaerer \& de Barros(2009)]{scha09} Schaerer, D., \& de Barros, S.\ 2009, \aap, 502, 423 
\bibitem[Schechter(1976)]{sche76} Schechter, P.\ 1976, \apj, 203, 297 
\bibitem[Schlafly \& Finkbeiner(2011)]{schl11} Schlafly, E.~F., \& Finkbeiner, D.~P.\ 2011, \apj, 737, 103 
\bibitem[Schmidt et al.(2014)]{schm14} Schmidt, K.~B., Treu, T., Trenti, M., et al.\ 2014, \apj, 786, 57 
\bibitem[Schulze et al.(2015)]{schu15} Schulze, S., Chapman, R., Hjorth, J., et al.\ 2015, \apj, 808, 73 
\bibitem[Stanway et al.(2016)]{sta15} Stanway, E.~R., Eldridge, J.~J., \& Becker, G.~D.\ 2016, \mnras, 456, 485 
\bibitem[Stark et al.(2015)]{stark15} Stark, D.~P., Richard, J., Charlot, S., et al.\ 2015, \mnras, 450, 1846 
\bibitem[Starling et al.(2013)]{sta13} Starling, R.~L.~C., Willingale, R., Tanvir, N.~R., et al.\ 2013, \mnras, 431, 3159 
\bibitem[Steidel et al.(1996)]{ste96} Steidel, C.~C., Giavalisco, M., Pettini, M., Dickinson, M., \& Adelberger, K.~L.\ 1996, \apjl, 462, L17
\bibitem[Svensson et al.(2010)]{sve10} Svensson, K.~M., Levan, A.~J., Tanvir, N.~R., Fruchter, A.~S., \& Strolger, L.-G.\ 2010, \mnras, 405, 57 
\bibitem[Tanvir et al.(2009)]{tan09} Tanvir, N.~R., Fox, D.~B., Levan, A.~J., et al.\ 2009, \nat, 461, 1254 
\bibitem[Tanvir et al.(2012)]{tan12} Tanvir, N.~R., Levan, A.~J., Fruchter, A.~S., et al.\ 2012, \apj, 754, 46 
\bibitem[Th{\"o}ne et al.(2013)]{tho13} Th{\"o}ne, C.~C., Fynbo, J.~P.~U., Goldoni, P., et al.\ 2013, \mnras, 428, 3590   
\bibitem[Totani et al.(2006)]{tot06} Totani, T., Kawai, N., Kosugi, G., et al.\ 2006, \pasj, 58, 485
\bibitem[Totani et al.(2014)]{tot14} Totani, T., Aoki, K., Hattori, T., et al.\ 2014, \pasj, 66, 63 
\bibitem[Trenti et al.(2012)]{tre12} Trenti, M., Perna, R., Levesque, E.~M., Shull, J.~M., \& Stocke, J.~T.\ 2012, \apjl, 749, L38 
\bibitem[Trenti et al.(2015)]{tre15} Trenti, M., Perna, R., \& Jimenez, R.\ 2015, \apj, 802, 103
\bibitem[Vergani et al.(2015)]{ver15} Vergani, S.~D., Salvaterra, R., Japelj, J., et al.\ 2015, \aap, 581, A102  
\bibitem[Wilkins et al.(2013)]{wil13} Wilkins, S.~M., Bunker, A., Coulton, W., et al.\ 2013, \mnras, 430, 2885 
\bibitem[Zafar et al.(2010)]{zaf10} Zafar, T., Watson, D.~J., Malesani, D., et al.\ 2010, \aap, 515, A94
\bibitem[Zafar et al.(2011)]{zaf11} Zafar, T., Watson, D.~J., Tanvir, N.~R., et al.\ 2011, \apj, 735, 2  
\end{thebibliography}
\end{document}